\documentclass[aps,prl,preprint,tightenlines,superscriptaddress,showpacs,byrevtex]{revtex4}
\usepackage{amsmath2000}
\usepackage{graphics}
\usepackage{dcolumn}

\begin{document}

\vspace*{-3\baselineskip}
\resizebox{!}{3cm}{\includegraphics{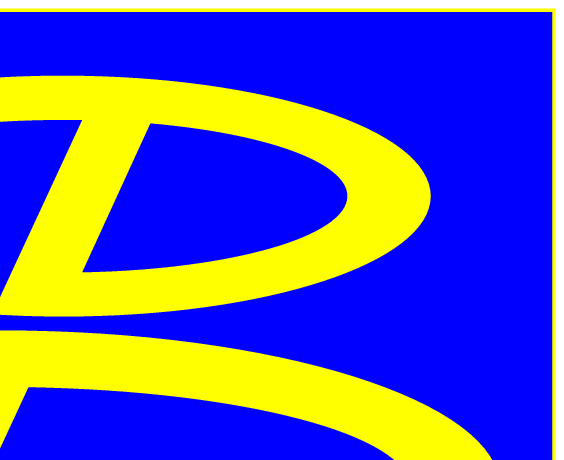}}

\preprint{KEK Preprint 2001-165}
\preprint{Belle Preprint 2002-2}

\title{\boldmath Precise Measurement of $B$ Meson Lifetimes \\
  with Hadronic Decay Final States}

\affiliation{Aomori University, Aomori}
\affiliation{Budker Institute of Nuclear Physics, Novosibirsk}
\affiliation{Chiba University, Chiba}
\affiliation{Chuo University, Tokyo}
\affiliation{University of Cincinnati, Cincinnati OH}
\affiliation{University of Frankfurt, Frankfurt}
\affiliation{Gyeongsang National University, Chinju}
\affiliation{University of Hawaii, Honolulu HI}
\affiliation{High Energy Accelerator Research Organization (KEK), Tsukuba}
\affiliation{Hiroshima Institute of Technology, Hiroshima}
\affiliation{Institute for Cosmic Ray Research, University of Tokyo, Tokyo}
\affiliation{Institute of High Energy Physics, Chinese Academy of Sciences, Beijing}
\affiliation{Institute of High Energy Physics, Vienna}
\affiliation{Institute for Theoretical and Experimental Physics, Moscow}
\affiliation{J.~Stefan Institute, Ljubljana}
\affiliation{Kanagawa University, Yokohama}
\affiliation{Korea University, Seoul}
\affiliation{Kyoto University, Kyoto}
\affiliation{Kyungpook National University, Taegu}
\affiliation{IPHE, University of Lausanne, Lausanne}
\affiliation{University of Ljubljana, Ljubljana}
\affiliation{University of Maribor, Maribor}
\affiliation{University of Melbourne, Victoria}
\affiliation{Nagoya University, Nagoya}
\affiliation{Nara Women's University, Nara}
\affiliation{National Kaohsiung Normal University, Kaohsiung}
\affiliation{National Lien-Ho Institute of Technology, Miao Li}
\affiliation{National Taiwan University, Taipei}
\affiliation{H.~Niewodniczanski Institute of Nuclear Physics, Krakow}
\affiliation{Nihon Dental College, Niigata}
\affiliation{Niigata University, Niigata}
\affiliation{Osaka City University, Osaka}
\affiliation{Osaka University, Osaka}
\affiliation{Panjab University, Chandigarh}
\affiliation{Peking University, Beijing}
\affiliation{Princeton University, Princeton NJ}
\affiliation{Seoul National University, Seoul}
\affiliation{Sungkyunkwan University, Suwon}
\affiliation{University of Sydney, Sydney NSW}
\affiliation{Tata Institute for Fundamental Research, Bombay}
\affiliation{Toho University, Funabashi}
\affiliation{Tohoku Gakuin University, Tagajo}
\affiliation{Tohoku University, Sendai}
\affiliation{University of Tokyo, Tokyo}
\affiliation{Tokyo Institute of Technology, Tokyo}
\affiliation{Tokyo Metropolitan University, Tokyo}
\affiliation{Tokyo University of Agriculture and Technology, Tokyo}
\affiliation{Toyama National College of Maritime Technology, Toyama}
\affiliation{University of Tsukuba, Tsukuba}
\affiliation{Utkal University, Bhubaneswer}
\affiliation{Virginia Polytechnic Institute and State University, Blacksburg VA}
\affiliation{Yokkaichi University, Yokkaichi}
\affiliation{Yonsei University, Seoul}

\author{K.~Abe}               
\affiliation{High Energy Accelerator Research Organization (KEK), Tsukuba}
\author{K.~Abe}               
\affiliation{Tohoku Gakuin University, Tagajo}
\author{R.~Abe}               
\affiliation{Niigata University, Niigata}
\author{T.~Abe}               
\affiliation{Tohoku University, Sendai}
\author{I.~Adachi}            
\affiliation{High Energy Accelerator Research Organization (KEK), Tsukuba}
\author{Byoung~Sup~Ahn}       
\affiliation{Korea University, Seoul}
\author{H.~Aihara}            
\affiliation{University of Tokyo, Tokyo}
\author{M.~Akatsu}            
\affiliation{Nagoya University, Nagoya}
\author{Y.~Asano}             
\affiliation{University of Tsukuba, Tsukuba}
\author{T.~Aso}               
\affiliation{Toyama National College of Maritime Technology, Toyama}
% ...

\collaboration{Belle Collaboration}
\noaffiliation

\date{\today}

\begin{abstract}
  The lifetimes of the $\bzb$ and $\bm$ mesons are extracted from $\intl$
  of data collected with the Belle detector at the KEK $B$-factory.
  A fit to the decay length differences of neutral and charged $B$ meson
  pairs, measured in events where one of the $B$ mesons is fully
  reconstructed in several hadronic modes, yields $\taubz = \tbzresult$~ps,
  $\taubm = \tbmresult$~ps, and $\rbm = \rbmresult$.
\end{abstract}


\maketitle


Lifetime measurements of $\bzb$ and $\bm$ provide key input
to the determination of the Kobayashi-Maskawa matrix element
$|V_{cb}|$~\cite{CKM}.
Moreover, the ratio of the lifetimes is sensitive to effects beyond the
spectator model.
In the framework of heavy quark expansion, the $B$ lifetime ratio
is predicted to be equal to one, up to small corrections proportional to
$1/m_b^3$~\cite{o3mb}, where $m_b$ is the mass of the $b$ quark.
A recent theoretical study predicts $\rbm = 1.07 \pm 0.03$~\cite{r-th},
which agrees with the world average of measurements,
$\rbm = 1.073 \pm 0.027$~\cite{PDG}.
The \textit{BABAR} collaboration recently reported
$\rbm = 1.082 \pm 0.026 \pm 0.012$~\cite{BaBar}.
Accurate determinations of the $\bzb$ and $\bm$ lifetimes also provide
essential input to analyses of $CP$ violation and neutral $B$ mixing.

The analysis described here is based on a $\intl$ data sample,
which contains $31.3 \times 10^6$ $\bbbar$ pairs, collected with the Belle
detector~\cite{Belle} at the asymmetric-energy KEKB collider
(8.0~GeV electrons against 3.5~GeV positrons)~\cite{KEKB}.
Electron-positron annihilations produce $\ups$ mesons moving
along the electron beamline ($z$ axis) with $\bgu = 0.425$,
and decaying into $\bz\bzb$ or $\bp\bm$.
Since the $B$ mesons are nearly at rest in the $\ups$ center of mass system
(cms), $B$ lifetimes can be determined from the separation in $z$
between the two $B$ decay vertices.
The average separation is $c\tau_B\bgu \sim 200~\micron$,
where $\tau_B$ is the $B$ meson lifetime.
In this analysis, one of the $B$ decay vertices is determined from
a fully reconstructed $B$ meson, while the other is determined inclusively
using the rest of the tracks in the event.

The Belle detector consists of
a three-layer silicon vertex detector~(SVD),
a 50-layer central drift chamber~(CDC),
an array of 1188 aerogel Cherenkov counters~(ACC),
128 time-of-flight~(TOF) scintillation counters,
an electromagnetic calorimeter containing 8736 CsI(Tl) crystals~(ECL),
and 14 layers of 4.7-cm-thick iron plates interleaved with
a system of resistive plate counters~(KLM).
All subdetectors except the KLM are located inside
a 3.4~m diameter superconducting solenoid which provides
a 1.5~T magnetic field.
The impact parameter resolutions for charged tracks are measured to be
$\sigma_{xy}^2 = (19)^2 + (50/(p\beta\sin^{3/2}\theta))^2~\micron^2$
in the plane perpendicular to the $z$ axis and
$\sigma_{z}^2 = (36)^2 + (42/(p\beta\sin^{5/2}\theta))^2~\micron^2$
along the $z$ axis, where $\beta = pc/E$, $p$ and $E$ are
the momentum ($\Gevc$) and energy (GeV), and
$\theta$ is the polar angle from the $z$ axis.
The transverse momentum resolution is
$(\sigma_{\pt}/\pt)^2 = (0.0030/\beta)^2 + (0.0019\pt)^2$,
where $\pt$ is in $\Gevc$.

$\bzb$ and $\bm$ mesons are fully reconstructed in the following decay modes:
$\bzb \to \dplus\pim$, $\dstarp\pim$, $\dstarp\rhom$, $\jpsi\ks$,
$\jpsi\kstarzb$, $\bm \to \dz\pim$, and $\jpsi\km$~\cite{cc}.

Charged pion and kaon candidates are required to satisfy
selection criteria based on particle-identification likelihood functions
derived from specific ionization ($\dedx$) in the CDC,
time-of-flight, and the response of the ACC.
Electrons are identified using a combination of $\dedx$,
the response of the ACC, and the position, shape, and total energy
(\ie, $E/p$) of their associated ECL showers.
Muon candidates must penetrate the iron plates of the KLM
in a manner consistent with the muon hypothesis.

The reconstruction and selection criteria for $\ks \to \pip\pim$
and $\jpsi \to \ell^+\ell^-$ ($\ell = e, \mu$) are described
elsewhere~\cite{phi1}.
$\kstarzb$ candidates are reconstructed in the $\kstarzb \to \km\pip$
channel and required to have an invariant mass within $75~\Mevcsq$
of the average $\kstarzb$ mass.

Photon candidates are defined as isolated ECL clusters of more than
20~MeV that are not matched to any charged track.
$\piz$ candidates are reconstructed from pairs of photon candidates
with invariant masses between 124 and $146~\Mevcsq$.
A mass-constrained fit is performed to improve the $\piz$ momentum resolution.
A minimum $\piz$ momentum of $0.2~\Gevc$ is required.
$\rhom$ candidates are selected as $\pim\piz$ pairs having invariant masses
within $150~\Mevcsq$ of the average $\rhom$ mass.

Neutral and charged $D$ candidates are reconstructed
in the following channels:
$\dz \to \kpi$, $\kpipiz$, $\kpipipi$, and $\dplus \to \kpipi$.
$\dstarp \to \dz\pip$ candidates are formed by combining
a $\dz$ candidate with a slow and positively charged track,
for which no particle identification is required.

To reduce continuum background, a selection based on
the ratio of the second to zeroth Fox-Wolfram moments~\cite{FW}
and the angle between the thrust axes of the reconstructed and
associated $B$ mesons is applied mode by mode.

The decay vertices of the two $B$ mesons in each event are fitted
using tracks that are associated with at least one SVD hit
in the $r$-$\phi$ plane and at least two SVD hits in the $r$-$z$ plane
under the constraint that they are consistent with
the interaction point (IP) profile, smeared in the $r$-$\phi$ plane
by $21~\micron$, to account for the transverse $B$ decay length.
The IP profile is described as a three-dimensional Gaussian,
the parameters of which are determined in each run
(every 60,000 events in case of the mean position)
using hadronic events.
The size of the IP region is typically $\sigma_x \simeq 100~\micron$,
$\sigma_y \simeq 5~\micron$, and $\sigma_z \simeq 3$~mm,
where $x$ and $y$ denote horizontal and vertical directions, respectively.

In the case of a fully reconstructed $\bdx$ decay,
the $B$ decay point is obtained from the vertex position and
momentum vector of the reconstructed $D$ meson and
a track other than the slow $\pip$ candidate from $\dstarp$ decay.
For a fully reconstructed $\bpsix$ decay, the $B$ vertex is determined
using lepton tracks from the $\jpsi$.
The decay vertex of the associated $B$ meson is determined
from tracks not assigned to the fully reconstructed $B$ meson;
however, poorly reconstructed tracks (with a longitudinal position error
in excess of $500~\micron$) as well as tracks likely to come from
$\ks$ decays (forming the $\ks$ mass with another track,
or more than $500~\micron$ away from the fully reconstructed
$B$ vertex in the $r$-$\phi$ plane) are not used.

The quality of the fit is assessed only in the $z$ direction
(because of the tight IP constraint in the transverse plane),
using the following variable
$\xi \equiv (1/2n) \sum^n_i \left[ (z_\text{after}^i-z_\text{before}^i) /
  \varepsilon_\text{before}^i \right]^2$,
where $n$ is the number of tracks used in the fit,
$z_\text{before}^i$ and $z_\text{after}^i$ are the $z$ positions
of each track (at the closest approach to the origin)
before and after the vertex fit, respectively,
and $\varepsilon_\text{before}^i$ is the error of $z_\text{before}^i$.
A Monte Carlo (MC) study shows that $\xi$ does not depend
on the $B$ decay length.
We require $\xi < 100$ to eliminate poorly reconstructed vertices.
About 3\% of the fully reconstructed
and 1\% of the associated $B$ decay vertices are rejected.

\begin{figure}
  \resizebox{0.6\textwidth}{!}{\includegraphics{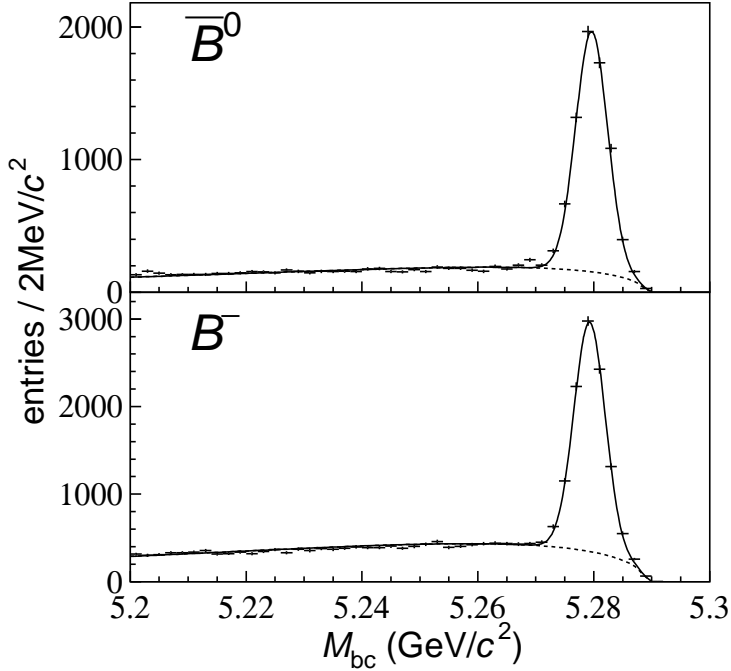}}%
  \caption{\label{fig:mbc}
    The beam-energy constrained mass distributions of the fully
    reconstructed $B$ candidates (neutral on top, charged on bottom).
    The dashed curves show the background contributions.}
\end{figure}

The proper-time difference between the fully-reconstructed
and the associated $B$ decays, $\dt \equiv \trec - \tasc$, is calculated
as $\dt =(\zrec - \zasc)/[c\bgu]$, where $\zrec$ and $\zasc$
are the $z$ coordinates of the fully-reconstructed and associated
$B$ decay vertices, respectively.
We reject a small fraction ($\sim$0.2\%) of the events by requiring
$|\dt| < 70$~ps ($\sim$45$\tau_B$).
The final event selection is based on requirements on
the energy difference $\dE \equiv \Eb - \Ebeam$ and
the beam-energy constrained mass $\mb \equiv \sqrt{(\Ebeam)^2-(\pb)^2}$,
where $\Ebeam$, $\Eb$, and $\pb$ are the beam energy, the energy,
and the momentum of the fully reconstructed $B$ candidate in the cms,
respectively.
If more than one fully reconstructed $B$ candidate is found
in the same event, the one with the best $\chisq$ for $\dE$ and $\mb$
(and the invariant mass of the $D$ candidate for the $\bdx$ case) is chosen.
For each channel, a square signal region is defined in the $\dE$-$\mb$
plane, corresponding to $\pm 3\sigma$ windows centered on the expected
means for $\dE$ and $\mb$.
A standard deviation is about $3~\Mevcsq$ for $\mb$ and
10--30~MeV for $\dE$ depending on the decay mode.
After all vertexing and selection requirements,
we find 7863 $\bzb$ and 12047 $\bm$ events in the $\dE$-$\mb$ signal region.
Figure~\ref{fig:mbc} shows the $\mb$ distributions for all
the $\bzb$ and $\bm$ candidates found in the $\dE$ signal region.

We extract lifetime information using an unbinned maximum likelihood fit
to the observed $\dt$ distributions.
We maximize the likelihood function $L = \prod_i P(\dt_i)$,
where $P(\dt)$ is the probability density function (PDF) for
the proper-time difference $\dt$ for each event and
the product is taken over all selected events.
The function $P(\dt)$, expressed as
\begin{eqnarray}
  P(\dt) &=& (1-\fol) \left[ \fsig \Psig(\dt)
    + (1-\fsig) \Pbg(\dt) \right] \nonumber \\*
  && + \fol \Pol(\dt) ,
\end{eqnarray}
contains contributions from the signal and the background
($\Psig$ and $\Pbg$), where $\Psig$ is described as the convolution
of a true PDF ($\calPsig$) with a resolution function ($\Rsig$)
and $\Pbg$ is expressed in a similar way:
\begin{equation}
  P_j(\dt) = \int_{-\infty}^{+\infty} \mathrm{d}(\dtp)
  \mathcal{P}_j(\dtp) R_j(\dt - \dtp) ,
\end{equation}
where $j=\text{sig}, \text{bkg}$.
To account for a small number of events that give
large $\dt$ in both the signal and background (outlier components),
we add a third component: $\Pol(\dt) = G(\dt; \sigol)$,
where $G(t; \sigma) \equiv \exp(-t^2/(2\sigma^2))/(\sqrt{2\pi}\sigma)$
is the Gaussian function.
The global fraction ($\fol$) and the width ($\sigol$)
are left as free parameters in the fit.

The signal purity ($\fsig$) is determined on an event-by-event basis
as a function of $\dE$ and $\mb$.
The two-dimensional distribution of these variables including
the sideband region is fitted with the sum of
a Gaussian signal function $\Fsig(\dE,\mb)$ and
a background function $\Fbg(\dE,\mb)$, represented as
an ARGUS background function~\cite{ARGUS} in $\mb$ and
a first-order polynomial in $\dE$.
The fraction of signal is then obtained as
$\fsig(\dE,\mb) = \Fsig/[\Fsig+\Fbg]$.
The true PDF for the signal is given by
\begin{equation}
  \calPsig(\dt) = \frac{1}{2\tau_B}\exp\left(-\frac{|\dt|}{\tau_B}\right) ,
\end{equation}
where $\tau_B$ is, depending on the reconstructed mode in the event,
either the $\bzb$ or the $\bm$ lifetime.
The resolution function of the signal is constructed as
the convolution of four different contributions:
the detector resolutions on $\zrec$ and $\zasc$ ($\Rrec$ and $\Rasc$),
an additional smearing on $\zasc$ due to the inclusion of tracks
which do not originate from the associated $B$ vertex ($\Rnp$),
mostly due to charm and $K_S$ decays,
and the kinematic approximation that the $B$ mesons are at rest
in the cms ($\Rk$).

For a vertex obtained from multiple tracks, $\Rrec$ and $\Rasc$
are parameterized as a single Gaussian function,
$G\left(t; (s_j^0 + s_j^1 \xi_j)\sigma_j\right)$ ($j=\text{rec}, \text{asc}$)
where $\sigma_j$ is the proper-time error estimated vertex-by-vertex
from the $z_j$ error, and $(s_j^0 + s_j^1 \xi_j)$ is a scale factor
based on the vertex fit quality $\xi$ defined above.
The parameters for the scale factor, $s_j^{0,1}$, are determined from
the lifetime fit. 
The typical value of the scale factor is
1.32 for $\Rrec$ and 1.20 for $\Rasc$.
For a vertex obtained from a single track~\cite{single_trk}, we use
$(1-\ftail) G\left(t; \smain \sigma_j \right)
+ \ftail G\left(t; \stail \sigma_j \right)$,
where $\smain$ and $\stail$ are global scaling factors.
The shape of $\Rnp$ is determined from MC data sample,
separately for $\bzb$ and $\bm$ events.
$\Rk$ is calculated analytically as a function of $\Eb$
and $\cosb$ from the kinematics of the $\ups$ two-body decay,
where $\thetab$ is the polar angle of the reconstructed $B$ in the cms.
The resulting $\dt$ resolution for the signal is $\sim$1.56~ps (rms).

The background PDF, $\Pbg(\dt)$, is modeled as a sum of exponential
and prompt components convoluted with
$\Rbg(\dt) = (1-\ftailbg) G\left(\dt; \smainbg\sqrt{\Srec^2+\Sasc^2}\right)
+ \ftailbg G\left(\dt; \stailbg\sqrt{\Srec^2+\Sasc^2}\right)$.
Different values are used for $\smainbg$, $\stailbg$, and $\ftailbg$
depending on whether both vertices are reconstructed
with multiple tracks or not.
The parameters for the background function $\Pbg$ are determined using
the $\dE$-$\mb$ sideband region for each decay mode.
A MC study shows that the fraction of prompt component
in the signal region is smaller (by $\sim$10--50\%
depending on the decay mode) than that in the sideband region.
We correct $\Pbg$ for this effect.

\begin{figure}
  \resizebox{0.6\textwidth}{!}{\includegraphics{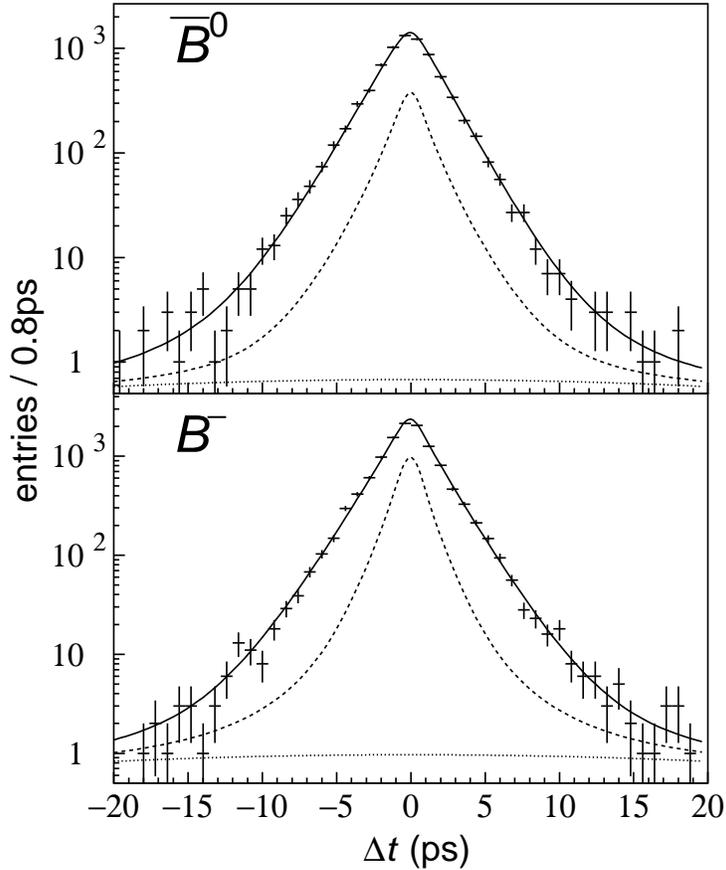}}%
  \caption{\label{fig:fit_result}
    The $\dt$ distributions of neutral (top) and charged (bottom)
    $B$ meson pairs, with fitted curves.  The dashed lines
    represent the sum of the background and outlier component,
    and the dotted lines represent the outlier component.}
\end{figure}

In the final fit to the events in the signal region,
we determine simultaneously 12 parameters:
the $\bzb$ and $\bm$ lifetimes, 7 parameters for $\Rrec$ and $\Rasc$,
and 3 parameters for the outlier component.
We find $\sigol = 36^{+5}_{-4}$~ps, and $\fol = (0.06^{+0.03}_{-0.02})$\%
or $(3.1 \pm 0.4)$\% (multiple- or single-track case).
The lifetime ratio, $r \equiv \rbm$, is obtained by repeating the
final fit after replacing $\taubm$ with $r\taubz$.
The fit yields $\taubz = \tbzresultmean \tbzresultstat$~ps,
$\taubm = \tbmresultmean \tbmresultstat$~ps,
and $\rbm = \rbmresultmean \rbmresultstat$.
Figure~\ref{fig:fit_result} shows the distributions of $\dt$
for $\bzb$ and $\bm$ events in the signal region
with the fitted curves superimposed.

The systematic errors are summarized in Table~\ref{tab:syserror}.
\begin{table}
  \caption{\label{tab:syserror}
    Summary of systematic errors for neutral and charged $B$ lifetimes,
    and their ratio.  The errors are combined in quadrature.}
  \begin{ruledtabular}
    \begin{tabular}{lddd}
                                & \multicolumn{1}{c}{$\taubz$ (ps)} & \multicolumn{1}{c}{$\taubm$ (ps)} & \multicolumn{1}{c}{$\rbm$} \\
      \hline
      IP constraint             & 0.004          & 0.003          & 0.001          \\
      Track selection           & 0.006          & 0.004          & 0.001          \\
      Vertex selection          & 0.003          & 0.002          & 0.002          \\
      $\dt$ range               & 0.003          & 0.002          & 0.001          \\
      $\dE$-$\mb$ signal region & 0.003          & 0.004          & 0.003          \\
      Signal fraction           & 0.001          & 0.001          & 0.001          \\
      $\Rsig$ parameterization  & 0.008          & 0.008          & \multicolumn{1}{r}{Cancels} \\
      $\Rnp$ parameters         & 0.006          & 0.004          & 0.006          \\
      Background shape          & 0.012          & 0.007          & 0.011          \\
      MC statistics             & 0.006          & 0.007          & 0.005          \\
      \hline
      Total                     & \tbzresultsyst & \tbmresultsyst & \rbmresultsyst \\
    \end{tabular}
  \end{ruledtabular}
\end{table}
The systematic error due to the IP constraint is estimated
by varying ($\pm 10~\micron$) the smearing used to account for
the transverse $B$ decay length.
The IP profile is determined using two different methods and
we find no change in the lifetime results.
Possible systematic effects due to the track quality selection
of the associated $B$ decay vertices are studied by varying
each criterion by 10\%.
The fit quality criterion for reconstructed vertices is varied
from $\xi < 50$ to $\xi < 200$.
We estimate the systematic uncertainty due to the maximum $|\dt|$
requirement by varying the $|\dt|$ range by $\pm 30$~ps and
taking the maximum excursion to be the systematic error.
We examine the uncertainty in the scale of $\dt$ arising from the
measurement error of the SVD sizes and thermal expansion during the
operation, and find that its contribution to the lifetimes
is negligibly small.
$\dE$-$\mb$ signal regions are varied by $\pm 10$~MeV for $\dE$
and $\pm 3~\Mevcsq$ for $\mb$.
The parameters determining $\fsig$ are varied by $\pm 1\sigma$
to estimate the associated systematic error.
The systematic error due to the modeling of $\Rsig$ is estimated
by comparing the results with different $\Rsig$ parameterizations.
The lifetime fit is repeated after varying the $\Rnp$ parameters
by $\pm 2\sigma$.
The lifetime dependence on the $B$ meson mass, which is the input for $\Rk$,
is measured by varying the mass by $\pm 1\sigma$ from the world average
value and found the differences to be negligible.
The systematic error due to the background shape is estimated
by varying its parameters by their errors.
The possible bias in the fitting procedure and the effect of
SVD alignment error are studied with MC samples;
we find no bias and include the MC statistics as a systematic error.

To conclude, we have presented new measurements of the $\bzb$ and $\bm$
meson lifetimes using $\intl$ of data collected with the Belle detector
near the $\ups$ energy.
Unbinned maximum likelihood fits to the distributions of the proper-time
difference between the two $B$ meson decays yield:
\begin{eqnarray*}
  \taubz &=& \tbzresult~\text{ps}, \\
  \taubm &=& \tbmresult~\text{ps}, \text{ and} \\
  \rbm &=& \rbmresult .
\end{eqnarray*}
These are currently the most precise measurements.
A value of unity for $\rbm$ is ruled out at a level
greater than $3\sigma$ and the measured value is consistent
with the theoretical prediction~\cite{r-th}.

\begin{acknowledgments}

We wish to thank the KEKB accelerator group.
We acknowledge support from the Ministry of Education,
Culture, Sports, Science, and Technology of Japan
and the Japan Society for the Promotion of Science;
the Australian Research Council
and the Australian Department of Industry, Science and Resources;
the National Science Foundation of China under contract No.~10175071;
the Department of Science and Technology of India;
the BK21 program of the Ministry of Education of Korea
and the CHEP SRC program of the Korea Science and Engineering Foundation;
the Polish State Committee for Scientific Research
under contract No.~2P03B 17017;
the Ministry of Science and Technology of the Russian Federation;
the Ministry of Education, Science and Sport of Slovenia;
the National Science Council and the Ministry of Education of Taiwan;
and the U.S.\ Department of Energy.

\end{acknowledgments}

\bibliography{blife_prep}

\begin{thebibliography}{12}
\bibitem[{cc()}]{cc}
\bibinfo{note}{Throughout this paper, when a decay mode is quoted the inclusion
  of charge conjugate mode is implied}.
\bibitem[{ARG()}]{ARGUS}
\bibinfo{note}{ARGUS Collaboration, H.~Albrecht \etal, \plb{241}{278}{1990};
  the ARGUS function is presented as $ax\sqrt{1-\left(x/\Ebeam\right)^2}
  \exp\left\{b\left[1-\left(x/\Ebeam\right)^2\right]\right\}$, where $a$ and
  $b$ are constants that are determined from the data.}
\bibitem[{sin()}]{single_trk}
\bibinfo{note}{The IP constraint enables us to determine a vertex even with a
  single track. The fraction of such vertices is about 1\% for $\zrec$ and 30\%
  for $\zasc$.}
\end{thebibliography}

\end{document}